\documentstyle[twocolumn,prl,aps,floats,psfig]{revtex} 
\begin{document}
\twocolumn[\hsize\textwidth\columnwidth\hsize\csname 
@twocolumnfalse\endcsname
\title{
Gauge factor of thick film resistors:
outcomes of the variable range hopping model} 
\author{C. Grimaldi, P. Ryser, S. Str\"assler} 
\address{D\'epartement de Microtechnique, IPM,
\'Ecole Polytechnique F\'ed\'erale de Lausanne,
CH-1015 Lausanne, Switzerland.}

\maketitle

\centerline \\

\begin{abstract}
Despite a large amount of data and numerous theoretical proposals,
the microscopic mechanism of transport in thick film resistors
remains unclear. However, recent low temperature measurements
point toward a possible variable range hopping mechanism of transport.
Here we examine how such a mechanism affects the gauge factor of thick
film resistors. We find that at sufficiently low temperatures $T$, for
which the resistivity follows the Mott's law 
$R(T)\sim \exp(T_0/T)^{1/4}$,
the gauge factor GF is proportional to $(T_0/T)^{1/4}$.
Moreover, the inclusion of Coulomb gap effects leads to 
${\rm GF}\sim (T_0'/T)^{1/2}$
at lower temperatures. In addition, we study a simple model which generalizes
the variable range hopping mechanism by taking 
into account the finite mean inter-grain spacing. Our results suggest
a possible experimental verification of the validity of the 
variable range hopping in thick film resistors.

\end{abstract}

\vskip 2pc ] 

\narrowtext
\centerline \\

\section{introduction}
\label{intro} 

Thick film resistors (TFRs) are composite materials in which metallic
grains (RuO$_2$, Bi$_2$Ru$_2$O$_7$, etc.) are embedded in an
insulating glassy matrix. The characteristic transport properties of 
these materials render the TFRs particularly suitable as
cryogenic thermometers and piezoresistive sensors.
Despite this successfull practical aspect, on the theoretical side
the situation is less satisfying. The microscopic mechanism
of transport in TFRs is in fact far from to be understood and several
theoretical models have been proposed, none of them however being
able to provide a completely satisfactory and definite answer.
 
Among the proposed transport mechanisms in TFRs, the most representative
ones are those based on tunneling within a network of
interconnected metallic clusters separated by thin glassy layers \cite{pike},
hopping between isolated metallic grains \cite{ping} and phonon-assisted
variable-range-hopping mechanism \cite{mott}.
The first model should be consistent with a highly dishomogenous 
structure in which the metallic phase is organized mainly in large
segments separated by thin layers of glass while the last two models
are more suitable for homogeneous dispersions of separated metallic
grains. The actual situation seems to lay somewhere in between these
two structures. 
Recent extended x-ray absorption fine structure experiments in fact have
revealed a bimodal distribution of the metallic particle sizes in
RuO$_2$-based TFRs\cite{meneghini}. According to these measurements,
the metallic phase is organized in large RuO$_2$ grains or clusters with
linear size ranging from $\sim 200$ \AA to 
$\sim 6000$ \AA and
much smaller RuO$_2$ grains with sizes of order $20$-$70$ \AA.
Where the actual current path takes place is still under debate.
However, It is plausible that transport in TFRs takes place mainly along
paths connecting the small metallic grains, the large RuO$_2$ clusters being
in fact too far away (a distance of order $100$ \AA as inferred from 
transmission electron microscopy \cite{tesi})
from each other for an electron to directly
hop from a large cluster to another large cluster.

TFRs have also a characteristic temperature dependence of the resitance $R(T)$.
At around room temperature, $R(T)$ has a minimum and it slightly increases at
higher $T$. At lower temperatures, $R(T)$ increases signalling a non-metallic
behavior. In this low-temperature region, $R(T)$ has been observed
to follow a $\exp(T_0/T)^x$ behavior with $x\simeq 1/4$ \cite{watson}
or $x\simeq 1/2$ \cite{schoepe}
depending on the room temperature resistance value of the samples.
Recent measurements \cite{prude2} have shown that $R(T)\sim\exp(T_0/T)^x$
with $x=1/4$ with a cross-over to $x=1/2$ as the temperature is lowered
for the most resistive samples.
Such kind of temperature dependence is typical of Mott
variable-range-hopping (VRH) mechanism of transport \cite{mott}
affected by Coulomb interaction between grains (Coulomb gap
effect \cite{es}).
Hence, the transport properties of TFRs at low temperatures
seem to be qualitatively described by ordinary VRH as other 
disordered systems like $n$-type CdSe \cite{aharony}.
However, the values of $T_0$ and $T_0'$ that best fit the experimental
data would lead to values of the optimal hopping length $R_h$ of the
same order of the localization length $\xi$, whereas the VRH 
mechanism should be observed only for $R_h\gg \xi$ \cite{prude2}.
The validity of the VRH mechanism is therefore not clear and additional 
informations are required to analyse the validity of the VRH approach in
describing transport in TFRs. 

Here we propose that a quantity which can be helpful and
easily measurable is the strain sensitivity, or gauge factor, 
and its temperature dependence.
The gauge factor (GF) relates the variation of the total resistance
$R(T)$ with an applied strain $\varepsilon$ and it is defined as follows:
\begin{equation}
\label{GF}
{\rm GF}=\frac{\delta R(T)}{\varepsilon R(T)} ,
\end{equation}
where $\delta R(T)$ is the variation of $R(T)$
under the applied strain $\varepsilon$.
TFRs have high values of GF, typically between $2$ and $35$ at room
temperature \cite{prude1}, and it is precisely this property that renders
these systems particularly suitable for piezoresistive sensor applications.
Here we are interested in studying how the VRH mechanism
affects GF and its temperature dependence. We find that {\it if the low 
temperature transport is due to the VRH mechanism} then the gauge factor
acquires a characteristic temperature dependence which could be
experimentally determined. 

In the next section we use the asymptotic regimes predicted by the
VRH model to extract qualitative behaviors for $GF$.
In Sec. \ref{simple}, a simple model is introduced with the aim of studying
the effect of finite inter-grain distances and to test the validity
of the qualitative behaviors of GF.

\section{Qualitative results }
\label{model}

In this section we analyze the transport properties of TFRs from
the point of view of VRH model. To this end, we must assume
drastic simplifications and we model therefore the TFRs structure as given
by an homogeneous dispersion of small metallic grains 
(of typical sizes $20$-$70$ $\mbox{\AA}$) in the
amorphous matrix and we neglect any possible effect of the larger metallic 
clusters on the bulk resistivity. As we discuss below, the assumption of
homogeneity probably leads to a strong under-estimation of the
room temperature GF value. However, instead of absolute values, here
we are mainly interested in relative
changes of GF as the temperature is varied.

An additional simplification we assume is to disregard the high temperature
rise of $R(T)$ beyond the temperature of minimum. Although the origin of this
feature is still debated, it should nevertheless been given by electron-phonon
intra-grain scattering, thermal expansion effects or maybe a combination of
both.  Within these approximations,
at sufficiently high temperatures transport is governed by tunneling
between adjacent grains which are separated by a mean distance $d$.
In this high temperature regime therefore
\begin{equation}
\label{rho1}
R(T)\equiv R_0\propto \exp(2d/\xi)
\end{equation}
where $\xi$ is the localization length and it is related to the barrier
potential $V$ between adjacent grains as $\xi=\hbar/\sqrt{2mV}$,
where $m$ is the electron mass. An applied strain modifies $d$ and
the resulting longitudinal GF becomes
\begin{equation}
\label{GF1}
{\rm GF}\equiv {\rm GF}_0=2d/\xi.
\end{equation} 
In the above expression it is implicitly assumed that the electrons
hop along paths mostly parallel to the direction in which strain is
applied. However, the actual microscopic paths are the result of 
hopping processes also along directions perpendicular to the applied
strain. In this way, the longitudinal gauge factor is 
something less than $2d/\xi$ and the amount of reduction depends
on specific material properties. However, as shown in appendix \ref{appa},
such a reduction can be estimated by
introducing a phenomenological parameter $\chi$ which measures
the percentage of hops along directions perpendicular to the strain,
leading to the following expression for the longitudinal gauge
factor:
\begin{equation}
\label{GF1bis}
{\rm GF}_0=\frac{1-\chi/(1-\nu)}{1+\chi}2d/\xi,
\end{equation} 
where $\nu$ is the Poisson ratio which varies between $0.2$ and $0.4$ for
typical TFRs \cite{prude1}. As shown in appendix \ref{appa}, 
$0< \chi <1/2$ and consequently $0.1 (2d/\xi) < {\rm GF}_0 < 2d/\xi$.
Since $\xi$ is of order of $d$, these values of GF$_0$
are much less than those measured in TFRs ($2< {\rm GF}_0 <35$).
We think that a strong enhancement of GF$_0$ could be achieved
when the assumption of homogeneity is relaxed. In fact, a strongly
non-homogeneous distribution of metal grains in the glassy matrix
could lead to a local strain distribution very different from the 
averaged one. In particular, the paths along which the current flows
can enter regions of concentrated strain leading therefore to an enhanced
total GF$_0$. Studies along this direction are currently under developement.

Having obtained the high temperature resistance and strain
sensitivity, we analize now the situation at low temperatures.
According to the VRH model, at sufficiently
low temperatures hopping to adjacent grains is no longer
favourable and $R(T)$ acquires a temperature dependence.
The question is whether the VRH mechanism
of transport affects the low temperature regime of
GF as well and if this variation can be experimentally tested.

In the VRH model, the transport properties are governed by
the probability $P_{ij}$ that an electron hops from 
grain $i$ to grain $j$ \cite{mott}:
\begin{equation}
\label{prob1}
P_{ij}\propto\exp\left[-2r_{ij}/\xi-\frac{E(r_{ij})}{K_{\rm B}T}\right],
\end{equation}
where $r_{ij}$ is the distance
between grains $i$ and $j$ and $E(r_{ij})$ is the energy threshold the 
electron experiences in hopping the distance $r_{ij}$. Assuming that
a finite fraction of the grains is charged, 
the energy $E(r_{ij})$ is made of two contributions \cite{aharony}:
\begin{equation}
\label{modelE1}
E(r_{ij})=\frac{\beta}{g r_{ij}^3}+\frac{\gamma e^2}{\kappa r_{ij}},
\end{equation}
where $g$ is the electron density of states, $e$ is the electron charge,
$\kappa$ is the dielectric constant, $\beta$ and $\gamma$ are dimensionless
constants. In the second right hand of equation (\ref{modelE1}), the first 
term describes the energy needed to hop to a grain a distance $r_{ij}$.
This term is proportional to $r_{ij}^{-3}$ because the probability an
electron at site $i$ has to find a site $j$ with energy much nearer
to its own scales as $r_{ij}^3$. 
The second term in Eq. (\ref{modelE1}) describes the Coulomb energy due
to finite charging of the grains, as it is expected by the fluctuations
in the energy levels in the grains.\cite{cuevas} This term is responsible
for the opening of the Coulomb gap.\cite{es}

At sufficiently low temperatures,  the resistance
is governed by $1/P_{\rm max}$ where $P_{\rm max}$ is the 
maximum of $P_{ij}$. 
Hence, the optimization of the exponential in the
hopping probability (\ref{prob1}) leads to a temperature dependence
of $R(T)$ characterized by two cross-over temperatures: 
\begin{equation}
\label{T0}
T_0=\frac{2048\beta}{27g \xi^3 K_B}\,, 
\end{equation}
which stems from the first term in the second hand side 
of Eq.(\ref{modelE1}), and 
\begin{equation}
\label{T01}
T_0'=\frac{8\gamma e^2}{\kappa \xi K_B}\, , 
\end{equation}
which is given by the Coulomb interaction. 
Usually $T_0'<T_0$ and $R(T)\sim\exp(T_0/T)^{1/4}$
for $T_0'< T < T_0$ while $R(T)\sim\exp(T_0'/T)^{1/2}$ for $T < T_0'$.

Depending on the particular temperature range, the strain sensitivity
of $R(T)$ is governed by the strain sensitivity of $T_0$ or $T_0'$.
Let us first focus on the temperature range in which 
$R(T)\sim\exp(T_0/T)^{1/4}$. An applied strain $\varepsilon$ leads to
following relation between the resistance variation $\delta R$ and
$\delta T_0$:
\begin{equation}
\label{var1}
\frac{\delta R}{R}=\frac{1}{4}\left(\frac{T_0}{T}\right)^{1/4}
\frac{\delta T_0}{T_0}.
\end{equation}
$T_0$ is inversely proportional to the electron density of states
$g$ which is proportional to the density of grains $n$. 
Assuming that an applied strain affects
the density of metallic grains leaving the volume of the grains 
unchanged \cite{note},
we obtain $\delta T_0/T_0=-\delta g/g=\varepsilon(1-2\nu)$ and the gauge
factor ${\rm GF}=\delta R/\varepsilon R$ becomes:
\begin{equation}
\label{GF2bis}
{\rm GF}=\frac{1-2\nu}{4}\left(\frac{T_0}{T}\right)^{1/4}.
\end{equation}
GF therefore increases as $T^{-1/4}$ as the temperature is lowered
provided $T$ is in the range for which $R(T)\sim\exp(T_0/T)^{1/4}$
holds true. 

By further lowering the temperature, transport becomes affected by the Coulomb
interaction and $R(T)$ crosses towards the $\sim\exp(T_0'/T)^{1/2}$
regime. Hence, under applied strain, $\delta R/R$ reduces to:
\begin{equation}
\label{var2}
\frac{\delta R}{R}=\frac{1}{2}\left(\frac{T_0'}{T}\right)^{1/2}
\frac{\delta T_0'}{T_0'}.
\end{equation}
The variation $\delta T_0'$ is driven by the strain dependence of
the total dielectric constant $\kappa$ and, in full generality, it can be
expressed as:
\begin{equation}
\label{var3}
\frac{\delta T_0'}{T_0'}=-\frac{\delta\kappa}{\kappa}=
\lambda\varepsilon(1-2\nu),
\end{equation}
where $\lambda$ is a dimensionless parameter and its value depends on 
the dielectric constants of the metallic and glassy phases.
In appendix \ref{appb} we provide an explicit expression of $\lambda$ based
on the Maxwell-Garnett formula for the dielectric response of
small metallic particles suspended in a dielectric.
By using (\ref{var2}) and (\ref{var3}), the gauge factor in the
Coulomb regime reduces to:
\begin{equation}
\label{GF3}
{\rm GF}=\frac{1-2\nu}{2}\lambda
\left(\frac{T_0'}{T}\right)^{1/2}.
\end{equation}
The Coulomb effect is therefore reflected in a cross-over from a
$T^{-1/4}$ to a $T^{-1/2}$ dependence of GF as the temperature
is sufficiently lowered. 
 
\section{Simple model for the GF temperature dependence}
\label{simple}

Summarizing the main results obtained in the previous section,
the temperature dependence of the resistance and the intrinsic 
GF as $T$ is lowered is characterized, within the VRH theory, by
three well distinguishable trends:
\begin{equation}
\label{trends}
\begin{array}{lll}
R\sim\exp(2d/\xi) & \rightarrow &{\rm GF}\sim2d/\xi \\
R\sim\exp\left(\frac{T_0}{T}\right)^{1/4} & \rightarrow & 
{\rm GF}\sim \left(\frac{T_0}{T}\right)^{1/4}\\
R\sim\exp\left(\frac{T_0'}{T}\right)^{1/2} & \rightarrow & 
{\rm GF}\sim \left(\frac{T_0'}{T}\right)^{1/2} \\
\end{array}
\end{equation}
Note that for each behavior listed above, GF is proportional
to $\ln\rho(T)$.
The last two behaviors originate from the optimization of the
inter-grain hopping probability $P_{ij}$ of Eq.(\ref{prob1}).
It is possible in fact to define a universal functional form for
$R(T)$ which contains both $\exp(T_0/T)^{1/4}$
and $\exp(T_0'/T)^{1/2}$ as limiting values and which fits well
transport data of Indium doped CdSe samples.~\cite{aharony}
However, for sufficiently high temperatures, the optimization
of $P_{ij}$ requires optimal hopping distances of order
or less than the mean distance $d$ between adjacent grains.
It would therefore be more correct to formulate the transport problem
in such a way that the value of $d$ is explicitly included.
Here we provide a simple version of such a formulation capable
of describing automatically the limiting behaviors
listed in Eq.(\ref{trends}).

A simple general equation for the conductance $G$ can be defined
as follows. A single hopping process between grains at distance $r_{ij}$
and with a hopping probability $P_{ij}$ can be regarded as a resistive
element with resistance $R_{ij}$ proportional to $1/P_{ij}$.
Hence, the total resistance can be constructed by interpreting
paths characterized by different hopping distances 
as resistors in a parallel geometry. The resulting conductance
should therefore be given by a weighted summation of 
the hopping probabilities $P_{ij}$ for all values of $r_{ij}\ge d$.
In this way, by employing a continuous approximation, we define the 
conductance $G(T)$  as:
\begin{equation}
\label{sigma1}
G(T)=a\int dr f(r)\exp\left[-2r/\xi-\frac{E(r)}{K_{\rm B}T}\right],
\end{equation}
where $E(r)$ is given in Eq.(\ref{modelE1}) and $a$ is a constant
introduced for dimensional purposes.
In the above equation, $f(r)$ is a weight function which takes
into account the different path geometries. 
For our purposes, the detailed structure of $f(r)$ is unnecessary
and we retain only its main feature which is given by
a lower cut-off representing
the mean distance $d$ between adjacent grains. Hence, we approximate 
$f(r)$ by the step function
$\theta(r-d)$. In this way, equation (\ref{sigma1}) reduces to:
\begin{equation}
\label{sigma2}
G(T)=a \int_d^{+\infty} 
dr \exp\left(-2r/\xi-\frac{E(r)}{K_{\rm B}T}\right).
\end{equation}
For temperatures larger than 
\begin{equation}
\label{tesse}
T_s=\frac{\xi E(d)}{2dK_{\rm B}} ,
\end{equation} 
equation (\ref{sigma2})
gives the proper high temperature limit (\ref{rho1}), while for $T\ll T_s$
the integral can be estimated by optimizing the exponential, leading
therefore to the VRH regime. 
Concerning the gauge factor resulting from (\ref{sigma2}), it should
be noted that an applied strain $\varepsilon$ affects both the interval of
integration and the energy threshold function $E(r)$ through the electron
density of states $g$ and the dielectric constant $\kappa$. Therefore,
by using the strain derivatives of $g$ and $\kappa$ introduced in
the previous section, GF can be expressed as the sum of three
different contributions, ${\rm GF}={\rm GF}_d+{\rm GF}_g+
{\rm GF}_{\kappa}$ where:
\begin{equation}
\label{Kd}
{\rm GF}_d=\frac{2d}{\xi}\left[\frac{1-\chi/(1-\nu)}{1+\chi}\right]
\frac{G(T)}{G_0}e^{-E(d)/K_{\rm B}T}
\end{equation}
\begin{equation}
\label{Kg}
{\rm GF}_g=\frac{1-2\nu}{K_{\rm B}T}\frac{a}{G(T)}
\int_d^{+\infty} dr \frac{\beta}{gr^3}
\exp\left(-2r/\xi-\frac{E(r)}{K_{\rm B}T}\right)
\end{equation}
\begin{equation}
\label{Kkappa}
{\rm GF}_{\kappa}=\frac{1-2\nu}{K_{\rm B}T}\frac{\lambda a}{G(T)}
\int_d^{+\infty} dr 
\frac{\gamma e^2}{\kappa r}\exp\left(-2r/\xi-\frac{E(r)}{K_{\rm B}T}\right)
\end{equation}
In equation (\ref{Kd}) $G_0=a\xi\exp(-2d/\xi)/2$ is the asymptotic
conductance in the high temperatures limit.

\begin{figure}[t]
\protect
\centerline{\psfig{figure=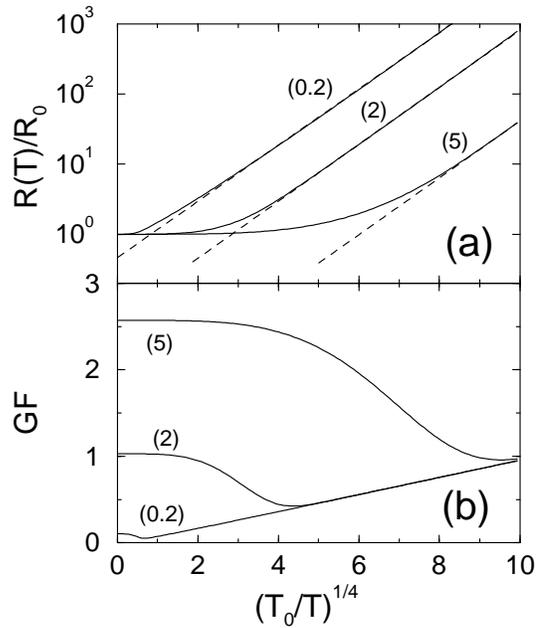,width=7cm}}
\caption{(a): temperature dependence of the resistance calculated from
Eq. (\protect\ref{sigma2}) without Coulomb interaction ($\gamma=0$) and for
the different values of $2d/\xi$ reported in the parentheses (solid lines).
The dashed lines fit the numerical data with $\exp(\widetilde{T}_0/T)^{1/4}$
where $\widetilde{T}_0\simeq 0.72 T_0$.
(b): corresponding temperature dependence of the gauge factor for
the same values of $2d/\xi$ as in Fig. 1a and for $\nu=0.3$ and
$\chi=1/4$.}
\label{GF_rho1}
\end{figure}

In Fig. 1a we plot the resistance $R(T)$ obtained numerically
from Eq.(\ref{sigma2}) in the absence of Coulomb repulsion ($\gamma=0$)
for $2 d/\xi=0.2$, $2$, and $5$. The resistance data are in
units of the high temperature limit $R_0=1/G_0$ and the
temperature is given in units of $T_0$ [see Eq.(\ref{T0})].
The curves show two
distinct regimes as a function of temperature due to the finite values
of the averaged inter-grain distance $d$. By using Eq.(\ref{tesse}), 
the cross-over between these two
regimes takes place roughly at $T\sim T_s=0.105 T_0/(2d/\xi)^4$, so that
as $2d/\xi$ increases the cross-over temprature $T_s$ is reduced.
For $T>T_s$ [$(T_0/T)^{1/4}<1.755 (2d/\xi)$]
the resistance shows
an activated regime while at sufficiently low temperatures
[$(T_0/T)^{1/4}>1.755 (2d/\xi)$] it becomes of VRH type. 
The latter behavior is signalled by the
dashed lines which fit the resistance with $R(T)/R_0\propto
\exp(\widetilde{T}_0/T)^{1/4}$, where $\widetilde{T}_0\simeq 0.75 T_0$.
The gauge factors GF associated to the resistance
curves of Fig. 1a are plotted in Fig. 1b for $\nu=0.3$ and $\chi=1/4$. 
A $2 d/\xi$ dependent
cross-over is obtained for GF as well. For temperatures higher than $T_s$, GF
is independent of $T$ and approaches the limit ${\rm GF}_0$ given in
Eq.(\ref{GF1bis}). On the contrary, by lowering $T$, GF acquires
a temperature dependence and for $T\ll T_s$ increases as
$1/T^{1/4}$ in agreement with Eq.(\ref{GF2bis}).
From the results of Fig. 1b, the $1/T^{1/4}$ dependence of the gauge factor
holds true in a wide range of temperatures only for quite small
values of ${\rm GF}_0$.  

\begin{figure}[t]
\protect
\centerline{\psfig{figure=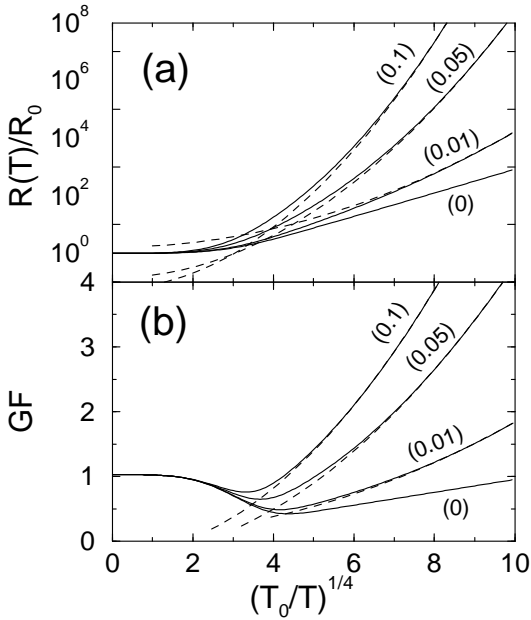,width=7cm}}
\caption{(a): temperature dependence of the resistance with the inclusion
of Coulomb interaction for $2d/\xi=2$ and $T_0'/T_0$ as reported
in the parentheses (solid lines).
The dashed lines fit the numerical data with $\exp(\widetilde{T}_0'/T)^{1/2}$
where $\widetilde{T}_0'\simeq 0.9 T_0'$.
(b): temperature dependence of the gauge factor for
the same values of $T_0'/T_0$ as in Fig. 1a, $\nu=0.3$, $\chi=1/4$,
and $\lambda=5$ (solid lines).
The dashed lines fit the low temperature regime of GF
with $(\widetilde{T}_0'/T)^{1/2}$.}
\label{GF_rho2}
\end{figure}

The effect of Coulomb repulsion on $R(T)$ and GF is displayed
in Fig. 2 where the case $2 d/\xi=2$ is investigated for different
values of the Coulomb parameter $T_0'$ defined in Eq.(\ref{T01}).
In Fig. 2a, for $T_0'/T_0\neq 0$ the resistance
approach the low temperature asymptotic regime $R(T)/R_0\propto
\exp(\widetilde{T}_0'/T)^{1/2}$ with $\widetilde{T}_0'\simeq 0.9 T_0'$ 
(dashed lines). The Coulomb effect is reflected also in the temperature
dependence of GF, Fig. 2b, where the curves have been calculated
by using $\lambda=1$, $\nu=0.3$ and $\chi=1/4$. 
The set of curves with $T_0'\neq 0$ show
the low temperature asymptotic behavior 
${\rm GF}\propto (\widetilde{T}_0'/T)^{1/2}$ with values of
$\widetilde{T}_0'$ somewhat larger than those obtained from 
the resistivity data ($\widetilde{T}_0'\simeq 1.05 T_0'$). Finally
the temperature dependence of GF for different values of $\lambda$
is plotted in Fig. 3 for the case $2d/\xi=2$ and 
$T_0'/T_0=0.05$.

\begin{figure}[t]
\protect
\centerline{\psfig{figure=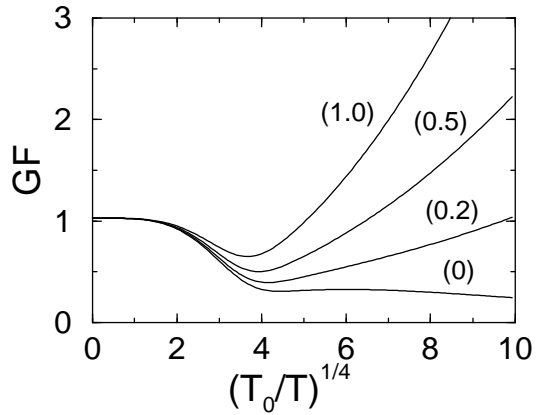,width=7cm}}
\caption{Gauge factor calculated for $T_0'/T_0=0.05$, $\nu=0.3$, $\chi=1/4$
and for different values of $\lambda$ (reported in the parentheses).}
\label{GF_rho3}
\end{figure}

\section{discussion and conclusions}
\label{concl}

The numerical solutions of equations (\ref{sigma2})-(\ref{Kkappa}) 
confirm the qualitative results given in the previous section,
{\it i.e.}, as the resistance crosses over the VRH regime, the
gauge factor acquires a temperature dependence which is
qualitatively summarized in Eq.(\ref{trends}). 

The model here presented is very simplified and susceptible
of various improvements. For example, from statistical arguments,
it would be certainly more correct to
evaluate the temperature dependence of $K(T)$ within the percolation
theory for transport \cite{perc}. This could be achieved by explicitly taking
into account in the evaluation of the critical path the lower 
cut-off $d$ for the hopping distance and the strain dependence of
$g$ and $\kappa$. As already stressed before,
an additional  simplifying assumption we have used is that of 
homogeneity. However
the structure of TFRs is very complex, being made by a glass
with dispersed metallic grains with bi-modal distribution of sizes.
In this situation,  under an applied strain,
the stress distribution inside the sample could strongly modify
GF as compared to the homogeneous case and to settle quantitatively 
this modification is certainly a very important issue.

A final open question regards the general applicability of the VRH concept
when electrons hop between grains of finite size. In fact, in reality,
the discrete electron energy levels of the grains are smeared by
finite life-time effects given by the electron-phonon and electron-electron
interactions together with the inter-grain tunneling coupling.
At sufficiently low temperatures, the life-time effects are small
and the discreteness of the energy levels requires phonon assisted
hopping for electron transport. However, at sufficiently high temperatures,
the energy smearing due to the finite electron life-time can be
of the order of the level spacing. In this situation it can be easily
shown that electrons can directly tunnel between grains without the
assistance of phonons. Of course, a crucial role is played by the
grain size which governs the energy level spacing. To our knowledge, the
problem of how the VRH between grains is affected by the inclusion of finite
life-time effects has never been addressed and its investigation is
certainly of the most valuable interest.

In conclusion, 
the results here reported provide
a clear prediction of the temperature dependence of the gauge factor in the
hypothesis that transport is well described by the classical VRH theory.
Hence, the simultaneous detection of the temperature 
dependences of the sheet resistance and the gauge factor can provide
an useful experimental tool to test the validity of the VRH mechanism
in TFRs.  We note that some experimental data show that GF for commercial
TFRs is largely independent of $T$ from room down to
cryogenics temperatures \cite{ferrero}. Our analysis would indicate
therefore that these experimental results are not compatible with
the VRH mechanism of transport. However, it is important to stress
that the VRH model predicts a strong temperature dependence of GF
only when the resistance crosses over a $\exp(T_0/T)^x$ type of
regime ($x=1/2$ or $1/4$), a test that has not been performed 
in Ref.\cite{ferrero}. Instead, a firm experimental test
requires measurements of the temperature dependences 
of both $R(T)$ and GF.

\acknowledgements
We would like to thank Th. M\"ader and M. Prudenziati
for interesting discussions.

\begin{appendix}
\section{}
\label{appa}

In this appendix we estimate the reduction of the gauge
factor due to the contributions of hopping processes perpendicular
to the direction on which strain is applied. To this end,
we first consider an operative definition of the gauge factor.
Let us consider a cantilever beam in the $x$-$y$ plane with axis 
lying on the $x$ direction on top of which is deposited the TFR. 
The thicknesses of the cantilever bean and of the TFR are measured
in the $z$ direction.
According to elasticity theory, bending of the cantilever beam
produces a finite strain $\varepsilon_{xx}=\varepsilon$ in the $x$ direction and
no strain in $y$ ($\varepsilon_{yy}=0$). Moreover,
if $\nu$ is the Poisson ratio, the strain of the TFR along the $z$ 
direction is $\varepsilon_{zz}=-\varepsilon\nu/(1-\nu)$.

To measure the longitudinal GF of the TFR, a potential difference must be
applied along the $x$ direction. If the microscopic paths are made
of discrete hopping processes exclusively along the direction of the
field, the longitudinal GF would be equal to $2d/\xi$ just as in
Eq.(\ref{rho1}). Conversely, the transversal GF would be zero because
in this case the field is applied in the $y$ direction for which
$\varepsilon_{yy}=0$.
Let us consider now the situation in which there is a finite percentage
$\chi$ of hops along directions perpendicular to that of the applied field.
The resulting resistance $\tilde{R}_x$, when the field is applied along
the $x$ direction, becomes $\tilde{R}_x=(1-\chi)R_x+\chi(R_y+R_z)$,
where $R_x$, $R_y$ and $R_z$ are resistances for hops in the $x$, $y$ and $z$ 
directions, respectively. Although $R_x$, $R_y$ and $R_z$ have the
same average value, they differ when strain is applied. In fact:
\begin{equation}
\label{straindep}
\frac{\delta R_x}{\varepsilon_{xx} R_x}=\frac{2d}{\xi}; \,
\frac{\delta R_y}{\varepsilon_{xx} R_y}=0; \,
\frac{\delta R_z}{\varepsilon_{xx} R_z}=-\frac{\nu}{1-\nu}\frac{2d}{\xi}.
\end{equation}
Hence the longitudinal gauge factor ${\rm GF}_L$ becomes:
\begin{eqnarray}
\label{lonGF1}
{\rm GF}_L \equiv\frac{\delta \tilde{R}_x}{\varepsilon_{xx} \tilde{R}_x} &=&
\frac{(1-\chi)\delta R_x+\chi \delta R_z}
{\varepsilon_{xx}[(1-\chi) R_x+\chi(R_y+R_z)]} \nonumber \\
&\simeq &\frac{1-\chi/(1-\nu)}{1+\chi}\frac{2d}{\xi} .
\end{eqnarray}
By following the same reasoning, the transversal gauge factor ${\rm GF}_T$
can be estimated by realizing that the resistance for a field applied
in the $y$ direction is $\tilde{R}_y=(1-\chi)R_y+\chi(R_x+R_z)$ and
consequentely:
\begin{eqnarray}
\label{perpGF1}
{\rm GF}_T \equiv\frac{\delta \tilde{R}_y}{\varepsilon_{xx} \tilde{R}_y} &=&
\frac{\chi (\delta R_x+\delta R_z)}
{\varepsilon_{xx}[(1-\chi) R_y+\chi(R_x+R_z)]} \nonumber \\
&\simeq &\frac{\chi[1-\nu /(1-\nu)]}{1+\chi}\frac{2d}{\xi} .
\end{eqnarray}
If the Poisson ratio $\nu$ is known, the phenomenological parameter
$\chi$ can be estimated by measuring the ratio between the
longitudinal and the transversal GFs which, according to (\ref{lonGF1})
and (\ref{perpGF1}), is given by:
\begin{equation}
\label{rapp}
\frac{{\rm GF}_L}{{\rm GF}_T}=\frac{1-\chi -\nu}{\chi (1-2\nu)}.
\end{equation}
Experimentally this ratio is found to be larger than the unity and
consequently the above expression leads to $\chi < 1/2$.

\section{}
\label{appb}

Here, we evaluate the strain dependence of the dielectric constant
for TFRs. We model the structure as given by spherical metallic particles
dispersed in a glassy matrix. The metallic and glassy phases occupy 
volumes $V_{\rm m}$ and $V_{\rm g}$, respectively, 
so that the volume fraction of
the metallic phase is $\phi=V_{\rm m}/(V_{\rm m}+V_{\rm g})\simeq
V_{\rm m}/V_{\rm g}$ for low metallic concentrations. Within the dipole
approximation, the total dielectric constant $\kappa$ is given by the
Maxwell-Garnett formula \cite{felderhof,book1}:
\begin{equation}
\label{m-g}
\kappa=\kappa_{\rm g}\left[1+\frac{3\phi}{1-\phi}\right],
\end{equation}
where $\kappa_{\rm g}$ is the dielectric constant of the glass. For 
sufficiently small values of $\phi$, the variation $\delta \kappa$
given by an applied strain $\varepsilon$ is:
\begin{equation}
\label{m-g2}
\delta\kappa=\kappa \frac{\delta\kappa_{\rm g}}{\kappa_{\rm g}}-
3\kappa_{\rm g} \phi\varepsilon (1-2\nu),
\end{equation}
where we have considered the metallic particles as perfectly rigid
bodies compared to the glass so that 
$\delta \phi=-\phi\varepsilon (1-2\nu)$ \cite{note}.

To find the variation $\delta\kappa_{\rm g}$ we should know an explicit
expression for $\kappa_{\rm g}$ which is however a difficult problem.
Nevertheless, experiments suggest that dielectric response of glasses
is under several aspects quite similar to that of ionic solids and, 
in particular, it has been shown that
the variation of the glass polarizability with pressure resembles
that of the ionic crystalline compounds \cite{book1}. 
By starting with this latter
observation we could argue that, for our purposes, the dielectric
constant for simple glasses is roughly given by:
\begin{equation}
\label{ionic1}
\kappa_{\rm g}=\frac{1+8\pi \alpha/3\Omega}
{1-4\pi \alpha/3\Omega},
\end{equation}
where $\Omega$ has dimension of volume and $\alpha\simeq 1/K$
is the ionic polarization with $K$ being the elastic constant. 
An expression similar to (\ref{ionic1})
is also recovered in theoretical analyses based on coarse-graining approach
\cite{book2} (there, $\Omega$ is a course-graining volume). 
An applied strain modifies $\kappa_{\rm g}$ via the volume $\Omega$
and the elastic constant $K$. The strain dependence of the latter
quantity can be guessed by observing that $K=V_{\rm ion}''(r_0)$ where
$V_{\rm ion}(r)$ is the interaction energy of two ions a distance $r$
apart and $r_0$ is the equilibrium distance. A pedagogical model for
$V(r)$ is $V(r)=-a/r+\exp(-br)+{\rm const.}$ \cite{bransden}, 
so that one obtains
$K=V_{\rm ion}''(r_0)\sim 1/r_0^2$. This result suggests that
$\alpha/\Omega$ scales as $1/[{\rm distance}]$ and consequently
from Eq.(\ref{ionic1}) one obtains in a straightforward way:
\begin{equation}
\label{ionic2}
\delta\kappa_{\rm g}\simeq -\frac{(\kappa_{\rm g}-1)
(\kappa_{\rm g}+2)}{3}\varepsilon (1-2\nu).
\end{equation}
The above expression plugged into Eq.(\ref {m-g2}) leads finally to:
\begin{eqnarray}
\label{m-g3}
\frac{\delta\kappa}{\kappa}&\simeq& -\left[\frac{(\kappa_{\rm g}-1)
(\kappa_{\rm g}+2)}{3\kappa_{\rm g}}
+3\frac{\kappa_{\rm g}}{\kappa} \phi\right]\varepsilon (1-2\nu) \nonumber \\
&\equiv& -\lambda\varepsilon (1-2\nu),
\end{eqnarray}
where the last equality defines the dimensionless quantity $\lambda$
used in the main text.

\end{appendix}


\begin{references}
\bibitem{note}
We are not aware of measurements reporting the Young modulus $E$
of bulk RuO$_2$. However, for metallic ruthenium the Young modulus
is quite large ($E\simeq 450$ GPa) and it is expected that RuO$_2$
has a value of $E$ of the same order. Since typical glasses have $E$ of order
$70$-$100$ GPa, it is a satisfying first approximation to consider the RuO$_2$
grains as perfectly rigid compared to the embedding glass.
\end{references}
\end{document}